\documentstyle[preprint,eqsecnum,aps]{revtex}

\begin{document}
\draft
\title{Breakdown of Migdal's theorem and intensity of electron-phonon coupling in
high-$T_c$ superconductors}
\author{G.A. Ummarino and R.S. Gonnelli}
\address{$INFM-$Dipartimento di Fisica, Politecnico di Torino, 10129 Torino, Italy}
\address{}
\date{October 9, 1997}
\maketitle

\begin{abstract}
In this article we quantify the possible effects of the breakdown of
Migdal's theorem on the electron-phonon coupling constant $\lambda $, on the
critical temperature $T_c$ and on the superconducting gap $\Delta $ by
examining different kinds of superconducting materials either with low and
high critical temperature. We use the theoretical approach developed by
Grimaldi, Pietronero and Str\"assler [PRB {\bf 52}, 10516 \& 10530, 1995] on
experimental data taken both from literature and from our recent
break-junction tunneling experiments in Bi$_2$Sr$_2$Ca Cu$_2$O$_{8+x}$. The
results show that a large violation of the Migdal's theorem (as in Bi$_2$Sr$%
_2$Ca Cu$_2$O$_{8+x}$) yields to a large increase of $\lambda $ and, in a
first approximation, to a large but different increase of $T_c$ and $\Delta $%
. The same theory gives no modifications when applied to low-$T_c$
conventional superconductors.
\end{abstract}


\preprint{HEP/123-qed}

\narrowtext

The high-$T_c$ superconductors (HTS), included the fullerenes, are all
characterized by a very low Fermi energy, here $E_F$, that is comparable
with the Debye phonon frequency: this involves the breakdown of the Migdal's
theorem for the electron-phonon (e-p) interaction \cite{ref42}, and requires
a generalization of the many-body theory of superconductivity.

Recently this generalization has been made \cite
{ref46,ref47,ref48,ref49,ref50} by using an extremely simplified model,
isotropous, tridimensional, mono-band and with the spectral function for the
e-p interaction represented by an appropriate Dirac function \cite{ref53}: $%
\alpha ^2F\left( \omega \right) =0.5\cdot \lambda \cdot \omega _0\cdot
\delta \left( \omega -\omega _0\right) $ where $\omega _0=$ $\omega _{\log
}=\exp \left( \frac 2\lambda \int_0^{+\infty }d\omega \frac{\alpha ^2F\left(
\omega \right) ^{ex}}\omega \ln \omega \right) $, $\lambda =2\int_0^{+\infty
}d\omega \frac{\alpha ^2F\left( \omega \right) ^{ex}}\omega $ and $\alpha
^2F\left( \omega \right) ^{ex}$ is the e-p spectral function experimentally
determined by quasiparticle tunneling measurements. In addition, a
perturbative scheme for what concerns the parameter $\left( \lambda \omega
_0/E_F\right) $ and a not perturbative one for $\lambda $ has been used. The
main result is that the vertex function shows a complex behaviour as
function of the phonon wavevector $q$ and frequency $\omega $ and, in
particular, it assumes positive values in correspondence of small $q$. The
region where $q$ is low can be favorite by the presence of electronic
correlations and by effects due to the electronic density of states \cite
{ref47}. This fact makes it possible a strong increase of the $T_c$ value,
in comparison with the results of conventional BCS theory. This approximate
vertex-corrected Migdal-Eliashberg theory, required by the breakdown of
Migdal's theorem and performed up to now only in {\it s} wave, at $T=T_c$
and with $\mu ^{*}=0$, gives rise to very interesting phenomena but, in
order to examine them in detail, it is necessary to solve numerically the
Eliashberg equations. This can be done by introducing a cutoff value for the
wavevector $q_c$, also useful in order to obtain some analytic results such
as the extension to this theory of the McMillan's formula for the
calculation of $T_c$ \cite{ref54,ref55}. As we already pointed out, the main
result is the strong $T_c$ increase which occurs together with a relatively
low $q_c$ and almost ordinary values for the coupling constant ($\lambda
\simeq 0.5-1$). A complex behaviour is present for the isotopic effect which
can be very high in correspondence of some parameters' regions, but can also
cancel when $\omega _0>E_F$. It is also interesting to note that
intermediate coupling constant values ($\lambda \simeq $ 1) can reproduce
properties which, in the standard Eliashberg theory, correspond to very high
values ($\lambda \simeq $ 3). The theory produces also effects on the
normal-state properties, which can draw away from the usual Fermi liquid
behaviour. At $T=T_c$, the new Eliashberg equations, that take into account
the first-order vertex corrections, are \cite{ref47,ref48}: 
\begin{eqnarray}
\Delta (\omega _n)Z(\omega _n) &=&\pi T_c\sum_{\omega _m}\frac{\omega _0^2}{%
\left( \omega _n-\omega _m\right) ^2+\omega _0^2}\frac{\Delta (\omega _n)}{%
\left| \omega _m\right| }\cdot  \nonumber  \label{1} \\
&&\lambda _\Delta (\omega _n,\omega _m,q_c,m,\lambda )\frac 2\pi \arctan
\left[ \frac{E_F}{Z(\omega _m)\left| \omega _m\right| }\right]  \eqnum{1}
\label{1}
\end{eqnarray}
\begin{eqnarray}
\left[ 1-Z(\omega _n)\right] &=&\frac{\pi T_c}{\omega _n}\sum_{\omega _m}%
\frac{\omega _0^2}{\left( \omega _n-\omega _m\right) ^2+\omega _0^2}\frac{%
\omega _m}{\left| \omega _m\right| }\cdot  \nonumber  \label{2} \\
&&\lambda _Z(\omega _n,\omega _m,q_c,m,\lambda )\frac 2\pi \arctan \left[ 
\frac{E_F}{Z(\omega _m)\left| \omega _m\right| }\right]  \eqnum{2}  \label{2}
\end{eqnarray}
where $\Delta (\omega _n)$ is the gap function, $Z(\omega _n)$ is the
renormalization function, $\omega _n=(2n-1)\pi k_BT$ with $n=0,\pm 1,\pm
2,...$ are the Matsubara frequencies and $m=\omega _0/E_F$ is a parameter
that represents a sort of indicator for the breakdown of the Migdal's
theorem. The functions $\lambda _\Delta $ and $\lambda _Z$ are defined as: 
\begin{eqnarray}
\lambda _\Delta (\omega _n,\omega _m,q_c,m,\lambda ) &=&\lambda \left[
1+\lambda P_c(\omega _n,\omega _m,q_c,m)\right]  \nonumber  \label{3} \\
&&+2\lambda ^2P_V(\omega _n,\omega _m,q_c,m)  \eqnum{3}  \label{3}
\end{eqnarray}

\begin{equation}
\lambda _Z(\omega _n,\omega _m,q_c,m,\lambda )=\lambda \left[ 1+\lambda
P_V(\omega _n,\omega _m,q_c,m)\right]  \eqnum{4}  \label{4}
\end{equation}
where $P_c(\omega _n,\omega _m,q_c,m)$ and $P_V(\omega _n,\omega _m,q_c,m)$
are cumbersome functions defined in the original papers\cite{ref47,ref48},
while $\lambda $ is the bare e-p coupling factor not renormalized by the
violation of the Migdal's theorem.

Following the usual procedure \cite{ref48,ref54,ref55}, it is possible to
simplify Eqs. 3 and 4 by calculating them at $\omega _n=0$ and $\omega
_m=\omega _0$ and then to obtain an approximate expression for $T_c$:

\begin{eqnarray}
T_c(q_c,m,\lambda ) &=&\frac{1.13\,\omega _0}{k_B\sqrt{e}(1+m)}\exp \left[ 
\frac m{2\left( 1+m\right) }\right] \cdot  \nonumber  \label{5} \\
&&\exp \left[ -\frac{1+\lambda _Z(q_c,m,\lambda )/\left( 1+m\right) }{%
\lambda _\Delta (q_c,m,\lambda )}\right] .  \eqnum{5}  \label{5}
\end{eqnarray}
In the limit $m\rightarrow 0,$ $\lambda _Z=\lambda _\Delta $ and this
formula coincides with the exact result obtained by Combescot \cite{ref56}
in the weak coupling regime.

Our original approach consists in numerically solving the system made by
Eqs. 4 and 5, with $T_c(q_c,m,\lambda )=T_c^{ex}$, and $\lambda
_Z(q_c,m,\lambda )=\lambda ^{ex}$, where $T_c^{ex}$ and $\lambda ^{ex}$ are
the experimental values for the critical temperature and for the e-p
coupling constant. In this way we obtain approximate values for $q_c$ and
the bare $\lambda $ value not renormalized by vertex corrections. We use the
experimental e-p spectral function $\alpha ^2F\left( \omega \right) ^{ex},$
determined from tunneling experiments, to calculate $\lambda ^{ex},$ $\omega
_0$ and the opportune Dirac's function $\alpha ^2F\left( \omega \right)
=0.5\cdot \lambda ^{ex}\cdot \omega _0\cdot \delta (\omega -\omega _0)$ that
has the property to give the critical temperature closest to the
experimental one in low-$T_c$ superconductors. Of course, the frequency $%
\omega _0$ also permits us to calculate the ratio $m$ which, as a
consequence, is not a free parameter of the model.

We applied this procedure to four different superconducting materials both
low-$T_c$ and high-$T_c$: Lanthanum, Bismuth, Ba$_{1-x}$K$_x$BiO$_3$ and Bi$%
_2$Sr$_2$CaCu$_2$O$_{8+x}$. The first two are conventional low-$T_c$
superconductors \cite{ref57} characterized by a gap of the order of 0.9-1.3
meV and $T_c\approx 5-6$ K. The main difference among them is the value of
the e-p coupling constant which is in the intermediate coupling regime for
La ($\lambda =0.98$) and in the very strong one for Bi ($\lambda =2.46$). Ba$%
_{1-x}$K$_x$BiO$_3$ (BKBO) is a well known ceramic superconductor with an
intermediate $T_c\simeq 24.5$ K and a cubic crystallographic structure.
Experiments have shown that the $\alpha ^2F(\omega )$ can be determined from
phonon anomalies in the quasiparticle tunneling conductance at $eV>\Delta $
and an intermediate $\lambda =1.23$ has been determined \cite{ref58}. In the
past years we have largely studied the high-$T_c$ superconductor Bi$_2$Sr$_2$%
CaCu$_2$O$_{8+x}$ (BSCCO) for the determination of the e-p spectral function 
\cite{ref15}. Very recently we obtained reproducible tunneling results on
optimally doped BSCCO single crystals with $T_c=93$ K \cite{ref59}. The
Eliashberg function $\alpha ^2F(\omega )$ was determined from break-junction
tunneling data and a good agreement was obtained at $eV>\Delta $ between the
experimental density of states and the theoretical one calculated by a
direct solution of the Eliashberg equations in presence of an
energy-dependent normal density of states. The coupling constant was
consistent with a very strong coupling regime ($\lambda =3.34$). Similar
results have been recently obtained by other groups \cite{ref17}. The
superconducting properties of the four materials are summarized in Table 1.
It is important to notice that we selected only materials characterized by
small or very small values of the Coulomb pseudopotential $\mu ^{*}\simeq 0$
(see Table 1), in order to be consistent with the approximate
vertex-corrected theory \cite{ref46,ref47,ref48} that has been written for $%
\mu ^{*}=0$.

Table 2 shows the logarithmic phonon energy $\omega _0$ calculated by the
already mentioned equation, the maximum phonon energy of the spectrum $%
\omega _{\max }$, the Fermi energy $E_F$ and the ratio $m=\omega _0/E_F$ for
the selected superconductors. Actually, in Bi, due to its amorphous
structure and the consequent particular shape of the spectral function at
low energy (i.e. for $\omega \rightarrow 0$ $\alpha ^2F\left( \omega \right)
\neq b\omega ^2$ with $b=const$) \cite{ref57}, we have used $\omega
_0=2A/\lambda $ where $A$ is the area of the spectral function. This $\omega
_0$ value, unlike $\omega _{\log }$, gives the correct critical temperature
in Bi. Due to the rather small value of $\omega _0$ (few meV) and the large
value of $E_F$ (some eV), the breakdown of Migdal's theorem is practically
absent in La and Bi ($m\sim 10^{-3}$ or less). Quite different situation is
present in BKBO and BSCCO where $\omega _0$ is of the order of 10-20 meV
while $E_F$ of the order of 100 meV. As a consequence, the violation of
Migdal's theorem is moderate in BKBO ($m\simeq 0.14$) and strong in BSCCO ($%
m\simeq 0.26$).

Figures 1 and 2 show the e-p spectral functions $\alpha ^2F(\omega )$ of the
selected superconducting materials taken from literature \cite{ref57} or
from recent tunneling experiments \cite{ref58,ref59}. The tiny vertical line
at abscissa $\omega _0$ in every graph indicates the position of the Dirac's
function which plays the role of $\alpha ^2F(\omega )$ in the approximate
vertex-corrected theory.

As already mentioned, by numerically solving the system of Eqs. 4 and 5 with

\begin{eqnarray*}
\lambda ^{ex} &=&\lambda _Z(q_c,\lambda ) \\
T_c^{ex} &=&T_c(q_c,\lambda )
\end{eqnarray*}
we can determine the cutoff value of the wavevector $q_c$ and the bare e-p
coupling constant $\lambda $.

It is important to remark that:

1) For the comparison with experimental values $\lambda ^{ex}$, we used $%
\lambda _Z(m,q_c,\lambda )$ and not $\lambda _\Delta (m,q_c,\lambda )$
because the renormalization effect on the coupling constant is present in
the normal state too;

2) The bare $\lambda $ values, determined by the above mentioned procedure,
are not the results of an exact numerical solution of the system of Eqs. 4
and 5 because such a solution generally does not exist. As a matter of fact
they are obtained by the average of the two values $\lambda _1$ and $\lambda
_2$ that fulfil the following conditions: $T_c(q_c,\lambda _2)=T_c^{ex}$ , $%
\lambda _Z(q_c,\lambda _1)=\lambda ^{ex}$ and $\Delta \lambda =\left|
\lambda _1-\lambda _2\right| $ is minimum ($\Delta \lambda _{BKBO}=0.08$ and 
$\Delta \lambda _{BSCCO}=0.3$). In practice we determine the smallest region
in the space of the parameters $q_c$ and $\lambda $ that is consistent with
an approximate numerical solution of the system. In the first three columns
of Table 3 the results of this procedure are shown for the selected
superconductors.

It is clear that for La and Bi, where the breakdown of Migdal's theorem is
absent, we obtain $\lambda =\lambda _Z=\lambda _\Delta =\lambda ^{ex}$ and
all the values of $q_c$ between 0.2-0.3 and 1 give an approximate solution
of the system. On the contrary, in BKBO the bare $\lambda $ is reduced of
the order of 20\% with respect to the experimental one and the solution is
possible only for an intermediate $q_c=0.6$. The largest renormalization is
present in BSCCO where the e-p coupling constant the superconductor would
have had in absence of effects due to the Migdal's theorem violation is $%
\lambda =1.85$, about 45\% less than the experimental value. In this case
the solution is possible only for a very small value of the wavevector $%
q_c=0.15$. The fourth column of Table 3 reports the values of the product $%
m\cdot \lambda $ which is the original parameter of the Migdal's expansion
and quantifies the amount of deviation from the standard Migdal-Eliashberg
theory.

After having calculated the bare $\lambda $, we determine a proper scaling
factor $\rho $ that, when applied to the $\alpha ^2F(\omega )^{ex}$, permits
to obtain a {\it scaled} e-p spectral function which gives a coupling
constant just equal to $\lambda $. The values of $\rho $ are shown in the
fifth column of Table 3. This scaled $\alpha ^2F(\omega )$ is then inserted
in a program for the direct solution of the standard Eliashberg equations at 
$T=0$ \cite{ref59} and with $\mu ^{*}=0$ to be consistent with the
vertex-corrected theory. By using this approach we determine a first
approximation for the values of the energy gap $\Delta _\lambda $ and the
critical temperature $T_{c\lambda }$ the superconductor would have had in
absence of the breakdown of Migdal's theorem. These values together with the
ratio $2\Delta _\lambda /k_BT_{c\lambda }$ are shown in the last three
columns of Table 3. Of course, due to the practical absence of lowest-order
vertex corrections, $\Delta _\lambda $ and $T_{c\lambda }$ of La and Bi are
exactly the same as the experimental values. In BKBO and BSCCO the effects
of these corrections on $\Delta $ and $T_c$ are very relevant. $\Delta
_\lambda $ and $T_{c\lambda }$ of BKBO are about 33\% and 28\% lower than
the corresponding experimental values (see Table 3) giving a ratio $2\Delta
_\lambda /k_BT_{c\lambda }=4$ that is consistent with an intermediate
coupling regime very similar to the La case. In BSCCO the reductions of $%
\Delta _\lambda $ and $T_{c\lambda }$ with respect to the experimental
values are about 48\% and 39\%, respectively, giving a renormalized gap of
12 meV and a renormalized critical temperature of 56 K. As a consequence of
the larger renormalization of $\Delta $ in comparison with $T_c$, the ratio $%
2\Delta _\lambda /k_BT_{c\lambda }$ reduces to 4.98, a value very close to
the ratio of Bi (see Table 3). We remark that some conventional low-$T_c$
very strong-coupling superconductors, like Pb$_{0.5}$Bi$_{0.5}$, have $%
\lambda $ and $2\Delta /k_BT_c$ values as large as 3 and 5.12, respectively 
\cite{ref18}.

Finally, in Figure 3, we show the dependence of the critical temperature of
BSCCO on $q_c$ and $m$ as determined from Eqs. 3 to 5. The light gray circle
on the contour map approximately shows the region of the parameters
corresponding to the solution of the system presented in Table 3. There are
two interesting observations: (i) it is clear from Fig. 3 and it was already
stressed in the original papers \cite{ref48} that, at the increase of the
breakdown of Migdal's theorem ($m\rightarrow 0.5$), $T_c$ increases only if $%
q_c$ is small. For $q_c$ of the order of 0.8-1, the increase of $m$ produces
a reduction of $T_c$; (ii) the action of removing the effects due to the
deviations from the standard Migdal-Eliashberg theory, that we have
tentatively done in the last part of the present paper, does not mean to
perform the limits $m\rightarrow 0$ and $q_c\rightarrow 1$ but $m\rightarrow
0$ while $q_c$ remains constant (in this case $q_c=0.15$). As a matter of
fact, $T_{c\lambda }\simeq 56$ K determined by the direct solution of the
Eliashberg equations is consistent with $m\simeq 10^{-3}$ and $q_c\simeq
0.15 $ as shown by the hollow circle in Fig.3. This guarantees a
self-consistency to our approach, since it has been calculated that $\lambda
\simeq 2$ is compatible with a ratio $T_c/\omega _0\simeq 0.2$ \cite{ref18}
which is the value we obtain by using $T_{c\lambda }$ for the critical
temperature.

In conclusion, the results of this paper show that the breakdown of Migdal's
theorem, certainly present at various degrees in high-$T_c$ superconductors,
yields to a relevant increase of the experimental e-p coupling constant $%
\lambda ^{ex}$ in comparison with the bare one. This increase appears
roughly proportional to the degree of violation of Migdal's theorem
expressed by $m\cdot \lambda $. By solving in direct way the standard
Eliashberg equations, we have shown that, in first approximation, $T_c$ and $%
\Delta $ are also increased by the effects of the approximate first-order
vertex corrections described in Refs. 2-6. The amount of such an increase is
greater for $\Delta $ than for $T_c$ thus leading to an amplified value of
the experimental ratio $2\Delta /k_BT_c$. In BSCCO, the renormalized values $%
\lambda =1.85$ and $2\Delta _\lambda /k_BT_{c\lambda }=4.98$ are quite
compatible with a conventional strong-coupling electron-phonon origin of
superconductivity.

We deeply acknowledge the useful discussions with O.V. Dolgov and C.
Grimaldi.

\begin{center}
{\bf TABLES}
\end{center}

Table 1: \\

\begin{tabular}{|cccccc|}
\hline
\multicolumn{1}{|c|}{} & \multicolumn{1}{c|}{$\lambda ^{ex}$} & 
\multicolumn{1}{c|}{$\mu _{ex}^{*}$} & \multicolumn{1}{c|}{$\Delta
^{ex}(meV) $} & \multicolumn{1}{c|}{$T_c^{ex}(K)$} & $2\Delta /k_BT_c$ \\ 
\hline
\multicolumn{1}{|c|}{$La$} & \multicolumn{1}{c|}{0.98} & \multicolumn{1}{c|}{
0.039} & \multicolumn{1}{c|}{0.89} & \multicolumn{1}{c|}{5.04} & 4.10 \\ 
\hline
\multicolumn{1}{|c|}{$Bi$} & \multicolumn{1}{c|}{2.46} & \multicolumn{1}{c|}{
0.091} & \multicolumn{1}{c|}{1.30} & \multicolumn{1}{c|}{6.11} & 4.90 \\ 
\hline
\multicolumn{1}{|c|}{$BKBO$} & \multicolumn{1}{c|}{1.23} & 
\multicolumn{1}{c|}{0.04} & \multicolumn{1}{c|}{4.5} & \multicolumn{1}{c|}{
24.5} & 4.64 \\ \hline
\multicolumn{1}{|c|}{$BSCCO$} & \multicolumn{1}{c|}{3.34} & 
\multicolumn{1}{c|}{0.0093} & \multicolumn{1}{c|}{23} & \multicolumn{1}{c|}{
93} & 5.74 \\ \hline
\end{tabular}
\\\\

Table 2: \\

\begin{tabular}{|ccccc|}
\hline
\multicolumn{1}{|c|}{} & \multicolumn{1}{c|}{$\omega _0(meV)$} & 
\multicolumn{1}{c|}{$\omega _{\max }(meV)$} & \multicolumn{1}{c|}{$E_F(meV)$}
& $m$ \\ \hline
\multicolumn{1}{|c|}{$La$} & \multicolumn{1}{c|}{4.49} & \multicolumn{1}{c|}{
15} & \multicolumn{1}{c|}{3300} & 0.0014 \\ \hline
\multicolumn{1}{|c|}{$Bi$} & \multicolumn{1}{c|}{2.87} & \multicolumn{1}{c|}{
14} & \multicolumn{1}{c|}{9900} & 0.0003 \\ \hline
\multicolumn{1}{|c|}{$BKBO$} & \multicolumn{1}{c|}{14.57} & 
\multicolumn{1}{c|}{62} & \multicolumn{1}{c|}{103} & 0.1415 \\ \hline
\multicolumn{1}{|c|}{$BSCCO$} & \multicolumn{1}{c|}{21.91} & 
\multicolumn{1}{c|}{90} & \multicolumn{1}{c|}{84} & 0.2608 \\ \hline
\end{tabular}
\\\\

Table 3: \\

\begin{tabular}{|ccccccccc|}
\hline
\multicolumn{1}{|c|}{} & \multicolumn{1}{c|}{$\lambda $} & 
\multicolumn{1}{c|}{$\lambda _\Delta $} & \multicolumn{1}{c|}{$q_c$} & 
\multicolumn{1}{c|}{$m\cdot \lambda $} & \multicolumn{1}{c|}{$\rho $} & 
\multicolumn{1}{c|}{$\Delta _\lambda (meV)$} & \multicolumn{1}{c|}{$%
T_{c\lambda }(K)$} & $2\Delta _\lambda /k_BT_{c\lambda }$ \\ \hline
\multicolumn{1}{|c|}{$La$} & \multicolumn{1}{c|}{0.98} & \multicolumn{1}{c|}{
0.98} & \multicolumn{1}{c|}{0.3-1} & \multicolumn{1}{c|}{0.0013} & 
\multicolumn{1}{c|}{1.00} & \multicolumn{1}{c|}{0.89} & \multicolumn{1}{c|}{
5.04} & 4.10 \\ \hline
\multicolumn{1}{|c|}{$Bi$} & \multicolumn{1}{c|}{2.46} & \multicolumn{1}{c|}{
2.46} & \multicolumn{1}{c|}{0.2-1} & \multicolumn{1}{c|}{0.0007} & 
\multicolumn{1}{c|}{1.00} & \multicolumn{1}{c|}{1.30} & \multicolumn{1}{c|}{
6.11} & 4.90 \\ \hline
\multicolumn{1}{|c|}{$BKBO$} & \multicolumn{1}{c|}{1.01} & 
\multicolumn{1}{c|}{1.4} & \multicolumn{1}{c|}{0.6} & \multicolumn{1}{c|}{
0.1429} & \multicolumn{1}{c|}{1.22} & \multicolumn{1}{c|}{3.05} & 
\multicolumn{1}{c|}{17.71} & 4.00 \\ \hline
\multicolumn{1}{|c|}{$BSCCO$} & \multicolumn{1}{c|}{1.85} & 
\multicolumn{1}{c|}{7.65} & \multicolumn{1}{c|}{0.15} & \multicolumn{1}{c|}{
0.4825} & \multicolumn{1}{c|}{1.81} & \multicolumn{1}{c|}{12.10} & 
\multicolumn{1}{c|}{56.34} & 4.98 \\ \hline
\end{tabular}
\\\\\newpage\ 

\begin{center}
{\bf FIGURE\ CAPTIONS}
\end{center}

Fig. 1 Experimental electron-phonon spectral functions $\alpha ^2F(\omega )$
of La (upper graph) and Bi (lower graph) taken from Ref. 11. The vertical
lines indicate the energy positions $\omega _0$ of the Dirac's functions
that play the role of the spectral functions in the approximate
vertex-corrected Migdal-Eliashberg theory.\\

Fig. 2 Experimental electron-phonon spectral functions $\alpha ^2F(\omega )$
of BKBO (upper graph, from Ref. 12) and BSCCO (lower graph, from Ref. 14).
Details are as in Fig.1.\\

Fig. 3 Contour map of the calculated critical temperature of BSCCO as a
function of $q_c$ and $m$. For details see the text.


\begin{references}
\bibitem{ref42}  A.B. Migdal, Sov. Phys. JETP {\bf 34,} 996 (1958).

\bibitem{ref46}  C. Grimaldi, L. Pietronero and S. Str\"assler, Phys. Rev.
Lett. {\bf 75,} 1158 (1995).

\bibitem{ref47}  L. Pietronero, S. Str\"assler and C. Grimaldi, Phys. Rev. B 
{\bf 52,} 10516 (1995).

\bibitem{ref48}  C. Grimaldi, L. Pietronero and S. Str\"assler, Phys. Rev. B 
{\bf 52,} 10530 (1995).

\bibitem{ref49}  G. Varelogiannis and L. Pietronero, Phys. Rev. B {\bf 52,}
15753 (1995).

\bibitem{ref50}  E. Cappelluti and L. Pietronero, Phys. Rev. B {\bf 53,} 932
(1996).

\bibitem{ref53}  G. Varelogiannis, Physica C {\bf 249,} 87 (1995).

\bibitem{ref54}  W.L. McMillan, Phys. Rev. {\bf 167,} 131 (1968).

\bibitem{ref55}  P.B. Allen and R.C. Dynes, Phys. Rev. B {\bf 12,} 905
(1975).

\bibitem{ref56}  R. Combescot, Phys. Rev. B {\bf 42,} 7810 (1990).

\bibitem{ref57}  E.L. Wolf, {\it Principles of electron tunneling
spectroscopy} (Oxford Univ. Press, New York, 1985).

\bibitem{ref58}  Q. Huang, J.F. Zasadzinski, N. Tralshawala, K.E. Gray, D.G.
Hinks, J.L. Peng and R.L. Greene, Nature {\bf 347,} 369 (1990).

\bibitem{ref15}  R.S. Gonnelli, D. Andreone, P. Samuely and S.I. Vedeneev,
in {\it Advances in high-temperature superconductivity}, eds. D.\ Andreone,
R.S. Gonnelli and E. Mezzetti (World Scientific, Singapore, 1992); R.S.
Gonnelli, S.I. Vedeneev, O.V. Dolgov and G.A. Ummarino, Physica C {\bf %
235-240,} 1861 (1994).

\bibitem{ref59}  R. S. Gonnelli, G. A. Ummarino and V. A. Stepanov,{\it \ }%
Physica C {\bf 275,} 162 (1997).

\bibitem{ref17}  D. Shimada {\it et al.}, Phys. Rev. B {\bf 51,} 16495
(1995).

\bibitem{ref18}  J.P. Carbotte, Rev. Mod. Phys. {\bf 62}, 1028 (1990).
\end{references}
\end{document}